\documentstyle[preprint,aps,prb]{revtex}

\begin{document}

\newcommand{\xik}{\chi_{\scriptscriptstyle \rm ALDA,\mu\nu}^{-1}}
\newcommand{\fik}{f^{\scriptscriptstyle \rm visc}_{\rm xc,\mu\nu}}
\newcommand{\bfp}{{\bf p}}
\newcommand{\bfk}{{\bf k}}
\newcommand{\bfq}{{\bf q}}
\newcommand{\dy}{\displaystyle}
\newcommand{\sinf}{\raisebox{-.7ex}{$\stackrel{<}{\sim}$}}
\newcommand{\ssup}{\raisebox{-.7ex}{$\stackrel{>}{\sim}$}}
\newcommand{\qp}{q_{||}}
\newcommand{\rp}{r_{||}}
\newcommand{\kp}{k_{||}}
\newcommand{\qqp}{{\bf q}_{||}}
\newcommand{\rrp}{{\bf r}_{||}}
\newcommand{\kkp}{{\bf k}_{||}}
\newcommand{\alda}{^{\scriptscriptstyle \rm ALDA}}
\newcommand{\dlda}{^{\scriptscriptstyle \rm DLDA}}
\newcommand{\vuc}{^{\scriptscriptstyle \rm VUC}}

\ifpreprintsty\else
\twocolumn[\hsize\textwidth%
\columnwidth\hsize\csname@twocolumnfalse\endcsname
\fi
\draft
\tightenlines
\preprint{ }
\title { Collective charge-density excitations of non-circular quantum dots 
in a magnetic field }
\author {C. A. Ullrich and G. Vignale}
\address
{Department of Physics,  University of Missouri, Columbia, Missouri 65211}
\date{\today}
\maketitle
\begin{abstract}
Recent photoabsorption measurements have revealed a rich fine structure
in the collective charge-density excitation spectrum
of few-electron quantum dots in the presence of magnetic fields.
We have performed systematic computational
studies of the far-infrared density response of quantum dots,
using time-dependent density-functional theory
in the linear regime and treating the dots
as two-dimensional disks. It turns out that
the main characteristics observed in the experiment
can be understood in terms of the
electronic shell structure of the quantum dots. However, new features arise
if a breaking of the circular symmetry of the dots is allowed, leading to
an improved description of the experimental results. 

\end{abstract}
\ifpreprintsty\else\vskip1pc]\fi
\narrowtext


\section{Introduction}

The electronic structure and dynamics of quantum dots have been a subject
of intense study in recent time.\cite{ashoori,hawrylak} 
Frequently one makes the simplifying
assumption that the electrons in quantum dots form a two-dimensional
system confined by a parabolic potential of circular symmetry. The
electronic ground state can then be found either through diagonalization
schemes or using mean-field approaches such as Hartree theory or
density-functional theory. The former treatment is computationally much
more demanding and has so far only been used to describe few-electron systems,
whereas the latter approach has been applied to electron numbers of order
100. In either case, however, one finds that the main features observed 
in the experiment (such as effects related to the electronic shell structure)
are well reproduced.

The model of a parabolic circular quantum dot is of course an idealization.
It has already been recognized in the early 90's that explaining certain
experimental features requires including some deviation from parabolicity
of the confining potential.
A first study of this kind has been performed by Broido {\em et al.},
\cite{broido} who considered the ground state and far-infrared (FIR)
response of up to 30 electrons confined in a circular quantum disk
of radius 100 nm. Their approach, as well as a similar one by
Gudmundsson and Gerhards,\cite{gudmundsson}  was 
based on a Hartree plus RPA description of the electrons.
Both studies showed that a correction to the confining potential 
towards the edge of the dot that makes it steeper than the parabolic potential
leads to a blue shift of the collective dipole modes that increases with the 
number $N$ of electrons in the dot. In addition, it was found in 
Ref. \onlinecite{gudmundsson} that one has to include    
a deviation from circular symmetry in order to explain the anticrossing behavior
in the FIR spectra observed in the experiments by Demel 
{\em et al}. \cite{demel}  These findings were supported
by an exact diagonalization study of quantum dot helium by Pfannkuche
and Gerhards.\cite{pfannkuche}
We finally mention a very recent study by Hirose and Wingreen \cite{hirose}
that uses spin-density-functional theory to describe
the electronic shell structure and calculate addition energies in elliptic dots.

The present work has been motivated by a recent series of experiments
on self-assembled InAs quantum dots performed in the group of 
J. P. Kotthaus in Munich. \cite{fricke,lorke,miller} 
The experiments use a combination of 
{\em in situ} capacitance spectroscopy and FIR absorption spectroscopy
to probe the ground state and collective charge-density excitations
as a function of the electron number per dot, $1 \le N \le 6$.
For $N=1,2$ one finds the well-known two-mode spectrum described by the
simple formula
\begin{equation}\label{1.1}
\omega_{\pm}=\sqrt{\omega_0^2 + \frac{\omega_c^2}{4}} \pm \frac{\omega_c}{2}\;,
\end{equation}
where $\omega_0$ is the characteristic frequency associated with a parabolic
confinement potential, and $\omega_c = e B/ m^* c$ is the cyclotron
frequency for particles of charge $e$ and mass $m^*$ in a magnetic field $B$.
However, for $N>2$ the authors of Refs. \onlinecite{fricke,lorke} detect a much 
richer FIR spectrum: the $\omega_+$ mode is observed to split up into
$\em three$ subpeaks (see Fig. 3 of Ref. \onlinecite{fricke}).

The authors argue within a simple picture of non-interacting particles
that the observed splitting of the $\omega_+$ mode
is caused by the non-parabolicity
of the confining potential of the quantum dots. Since for $N=3,4,5$ the
p-shell is only partly filled, the system can perform transitions of the
$\rm s \to p$ and $\rm p \to d$ type, which have slightly different energies,
in contrast with the strictly parabolic case. This simple explanation,
however, can only account for a two-fold splitting of the $\omega_+$ mode.
It is then further argued that the observed third (somewhat weaker) signal
is caused by effects related to electron-electron interaction.

It is the purpose of the present paper to provide a theoretical explanation
of the three-fold splitting of the $\omega_+$ mode observed in 
Ref. \onlinecite{fricke}. Our approach is based on density-functional
theory for the ground state and linear response in the presence of a magnetic
field. It will turn out in the following that
including electron-electron interaction effects alone is not sufficient.   
Instead, the presence of the third subpeak is explained by a combination
of the non-parabolicity {\em plus} breaking of the circular symmetry
of the confining potential.

The paper is organized as follows: in Section 2, we shall introduce
our model for the quantum dots, a two-dimensional disk,
and we shall present the theoretical methods for describing the electronic
ground state and dynamics. In Section 3, we then discuss our numerical results,
and Section 4 gives our conclusions.
\section{formal framework}\label{sec2}

\subsection{Ground-state calculation}\label{sec2.1}
The electronic ground state of a system in the presence of magnetic fields
is described by current-density-functional theory. The general formalism has
been developed by Vignale and Rasolt, \cite{vignale} and to date there
exist several applications to quantum dots. \cite{ferconi,spanish}
Here, we want to generalize the formalism, which originally had been developed
for circular quantum dots, to describe non-circular systems. 

In the following, we shall make our lives a little easier, especially in
view of the linear response calculations later on, and neglect
the dependence of the exchange-correlation (xc) energy $E_{\rm xc}$ on 
the current density ${\bf j}_{p\sigma}$. This dependence is known 
to cause only a small effect in the electronic ground state of quantum dots. 
The xc vector potential then drops out, and the Kohn-Sham equation
as derived in Ref. \onlinecite{vignale} becomes
\begin{eqnarray}\label{2.1.1}
\lefteqn{
\left\{ - \frac{ \nabla^2}{2m^*} - \frac{ie}{2m^*c} \:
({\bf A}_{\rm ext}({\bf r}) \cdot \nabla 
+ \nabla \cdot {\bf A}_{\rm ext}({\bf r}) )\right.}  \nonumber\\
&+& \left. \frac{e^2 A_{\rm ext}^2({\bf r})}{2m^*c^2}
+v_{\rm ext \sigma}({\bf r})+ v_{\rm H}({\bf r}) +  v_{\rm xc \sigma}({\bf r})
\right\} \psi_{j\sigma}({\bf r}) \nonumber\\
& = &
\epsilon_{j\sigma} \psi_{j\sigma}({\bf r}) \;, 
\end{eqnarray}
where as usual $v_{\rm ext \sigma}$ and ${\bf A}_{\rm ext}$ denote the
external scalar and vector potentials and $v_{\rm H}({\bf r})$ and
$v_{\rm xc \sigma}({\bf r})$ are the Hartree and xc potentials.
In the following, the basic assumption is that the quantum dots can be
treated as two-dimensional systems.   We then use polar coordinates and
write the external potential as a sum of the bare potential of
the quantum dot plus the Zeeman term:
\begin{equation}\label{2.1.2}
v_{\rm ext \sigma} (r,\varphi) = v_{\rm dot}(r,\varphi)
+ \frac{1}{2}\: g^* \mu_B B \sigma \;.
\end{equation}
Here, $\mu_B = e\hbar/2m$, $\sigma=\pm 1$,
and the specific form of $v_{\rm dot}(r,\varphi)$
will be discussed in section \ref{sec2.3}  below.
In turn, the external vector potential is given by
\begin{equation}\label{2.1.3}
{\bf A}_{\rm ext}(r,\varphi) = \frac{B}{2} \: r \: \hat{e}_\varphi \;,
\end{equation}
corresponding to a uniform magnetic field $B$ perpendicular to the dot.
Next, we expand the Kohn-Sham orbitals in polar coordinates as
\begin{equation}\label{2.1.4}
\psi_{j\sigma}({\bf r}) = \sum_l \phi_{jl\sigma}(r) \: e^{-il\varphi}\;,
\end{equation}
where the $ \phi_{jl\sigma}(r)$ are real.
The spin density is then given by
\begin{eqnarray}\label{2.1.5}
n_\sigma({\bf r}) = n_\sigma(r,\varphi)&=&
\sum_j f_{j\sigma} |\psi_{j\sigma}({\bf r})|^2 \nonumber\\
&=& \sum_{jll'}f_{j\sigma} \phi_{jl\sigma}(r) \phi_{jl'\sigma}(r)
\: e^{i(l-l')\varphi}\;,
\end{eqnarray}
where the $f_{j\sigma}$ denote occupation numbers of the orbitals.
In the following, all calculations will be performed at a small but finite
temperature, $T=4.2$ K, in order to avoid convergence problems at small
magnetic fields. The occupation numbers
are then given by the thermal distribution
\begin{equation}\label{2.1.6}
f_{j\sigma} = \frac{1}{1 + 
\exp\left({\frac{\epsilon_{j\sigma} - \mu}{k_B T}} \right)} \;,
\end{equation}
and the chemical potential $\mu$ is fixed through the relation
$\sum_{j\sigma} f_{j\sigma} = N$.
Next, the confining potential of the dot is expanded as
\begin{equation}\label{2.1.7}
v_{\rm dot}(r,\varphi) =
\sum_l v_{{\rm dot},l}(r) \: e^{-il\varphi} \;.
\end{equation}
Similar expansions are made for the Hartree and xc potential.
For the dot potential, the angular components 
\begin{equation}\label{2.1.8}
v_{{\rm dot},l}(r) = \frac{1}{2\pi} \int\limits_0^{2\pi} d\varphi\:
e^{il\varphi} \: v_{\rm dot}(r,\varphi) 
\end{equation}
in general have to be obtained through straightforward numerical integration,
and similarly for the xc potential (for the latter we use the local-density 
approximation in the parametrization of Tanatar and Ceperley \cite{tanatar}).
For the Hartree potential, one finds
\begin{equation}\label{2.1.9}
v_{{\rm H},l}(r) = 2\pi {e^*}^2\sum_\sigma \!  \sum_{jnn'\atop n-n'=l}
\int\limits_0^\infty r' dr'
\: \phi_{jn\sigma}(r') \phi_{jn'\sigma}(r') \: I_{l}(r,r') \;,
\end{equation}
where ${e^*}$ is the effective electronic charge, and
the $I_{l}(r,r')$ involve integrals over Bessel functions:
\begin{equation}\label{2.1.10}
I_{l}(r,r') = \int\limits_0^\infty \! dq \: J_{l}(qr) J_{l}(qr') \;.
\end{equation}
The $I_{l}(r,r')$ can be expressed  in terms of hypergeometric
functions that can be further reduced to complete elliptic integrals
(see Ref. \onlinecite{broido} for the case $l=1$).
Inserting everything into the Kohn-Sham equation, we finally obtain
\begin{eqnarray}\label{2.1.11}
\lefteqn{
\left\{ -\frac{1}{2m^*} \left( \frac{ d^2}{d r^2} +
\frac{1}{r} \frac{ d }{d r} - \frac{l^2}{r^2} \right)
- \frac{el B}{2m^*c} \right. } \nonumber\\
&+& \left. \frac{ e^2 B^2 r^2}{8m^*c^2} 
+ \frac{1}{2}\:g^* \mu_B B \sigma \right\}\phi_{jl\sigma}(r)
\nonumber \\
&+&   \sum_{k} \Big[
v_{{\rm dot},k}(r) +  v_{{\rm H},k}(r) +  v_{{\rm xc} \sigma,k}(r) \Big]
\phi_{j\,l-k \,\sigma}(r) \nonumber\\
& = & \epsilon_{j\sigma}\: \phi_{jl\sigma}(r) \;.
\end{eqnarray}
Eq. (\ref{2.1.11}) couples the angular components of the Kohn-Sham orbitals
(\ref{2.1.4}). In practice, we of course limit the expansions of the orbitals
and potentials to a 
finite number of components, $-L, \ldots, l, \ldots, L$, where $L$ of the
order 5 is sufficient to give convergence for the cases under study. 
Solution of the Kohn-Sham
equation (\ref{2.1.11}) is then accomplished by discretization on a 
logarithmic radial mesh with $N_{\rm grid}\sim 100$ grid points.

\subsection{Linear response}\label{sec2.2}
The FIR absorption spectra as measured in Refs. \onlinecite{fricke,lorke} are
proportional to the photoabsorption cross section
\begin{equation}\label{2.2.1}
\sigma(\omega) = \frac{4\pi\omega}{c}\: {\rm Im}\, \alpha(\omega)\;,
\end{equation}
where the dipole polarizability with respect to, say, the $x$-axis
\begin{equation}\label{2.2.2}
\alpha(\omega) = -\frac{2e}{E_0} \int d^3 r \: x \: n_1({\bf r},\omega)
\end{equation}
is obtained from the
linear density response $n_1({\bf r},\omega)$ of the quantum dots
to an external field of the form
\begin{equation}\label{2.2.3}
v_1({\bf r},\omega) = \frac{e E_0 x}{2} \;,
\end{equation}
where $E_0$ is the amplitude of the electric field strength.
In order to calculate $n_1({\bf r},\omega) = \sum_\sigma n_{1\sigma}
({\bf r},\omega)$, we have to solve the linear spin-density response equation,
which reads as follows:
\begin{eqnarray}\label{2.2.4}
\lefteqn{
n_{1\sigma}({\bf r},\omega) = 
\int\!d^3r'\: \chi_{\sigma}({\bf r},{\bf r}',\omega)
\bigg\{v_{1\sigma}({\bf r}',\omega)} \nonumber\\
&+&  \sum_\tau \int\!d^3r''
\left( \frac{{e^*}^2}{|{\bf r}' - {\bf r}''|}
+ f_{\rm xc \sigma \tau}({\bf r}',{\bf r}'',\omega) 
\right)n_{1\tau}({\bf r}'',\omega)
\bigg\} , \nonumber\\
&&
\end{eqnarray}
i.e. we have to solve two coupled integral equations for $n_{1\uparrow}$ and
$n_{1\downarrow}$. The Kohn-Sham response function 
\begin{equation}\label{2.2.5}
\chi_\sigma({\bf r},{\bf r}',\omega) = \sum_{j,k}^\infty (f_{k\sigma}
-f_{j\sigma}) \frac{\psi_{k\sigma}^*({\bf r})
\psi_{j\sigma}({\bf r})\psi_{j\sigma}^*({\bf r}')\psi_{k\sigma}({\bf r}')}
{\varepsilon_{k\sigma} - \varepsilon_{j\sigma} + \omega + i\eta} 
\end{equation}
is diagonal in the spins. Let us now expand the density response as
\begin{equation}\label{2.2.6}
n_{1\sigma}({\bf r},\omega) = \sum_{n=-N}^N
n_{1n\sigma}(r,\omega) e^{in\varphi} \;,
\end{equation}
where in practice of course $N \le L$. Inserting the form (\ref{2.1.4}) of
the Kohn-Sham orbitals, the response function is set up in the following way:
\begin{equation}\label{2.2.7}
\chi_\sigma({\bf r},{\bf r}',\omega) = 
\sum_{l,l'=-N}^N \chi_{ll' \sigma} (r,r',\omega) e^{il\varphi} e^{-il'\varphi'}
\;,
\end{equation}
where we define
\begin{equation}\label{2.2.8}
\chi_{ll' \sigma} (r,r',\omega) = 
\sum_{j,k}^\infty (f_{k\sigma} - f_{j\sigma})
\frac{ \Phi_{jkl}(r) \Phi_{jkl'}(r')}{\varepsilon_{k\sigma}
-\varepsilon_{j\sigma} + \omega + i\eta}
\end{equation}
with
\begin{equation}\label{2.2.9}
\Phi_{jkL}(r) = \sum_{l,l' \atop l'-l = L} \phi_{jl}(r)\phi_{kl'}(r) \;.
\end{equation}
For the xc kernel we use the adiabatic local-density approximation
(ALDA)
\begin{eqnarray}\label{2.2.10}
\lefteqn{
f_{\rm xc,\sigma \tau}({\bf r}',{\bf r}'',\omega) =
\left. \frac{d^2e_{\rm xc}}{dn_\sigma \, dn_\tau} \right|_{n_0({\bf r'})}
\delta({\bf r'} - {\bf r}'')} \nonumber\\
&=&
\delta({\bf r'} - {\bf r}'') \sum_{m'=-N}^N f_{{\rm xc},m'\sigma \tau}
(r') \: e^{im' \varphi'}\;,
\end{eqnarray}
where $e_{\rm xc}$ is the xc energy density of the homogeneous two-dimensional
electron gas, \cite{tanatar} and
\begin{equation}\label{2.2.11}
f_{{\rm xc},m'\sigma \tau}(r') =
\frac{1}{2\pi}\int\limits_0^{2\pi}d\varphi'\:
\left. \frac{d^2e_{\rm xc}}{dn_\sigma \, dn_\tau} \right|_{n_0({\bf r'})}
e^{-im' \varphi'} \;.
\end{equation}
In the ALDA, the xc kernel is frequency independent and real (for a recent
discussion of alternative expressions for $f_{\rm xc}$ see Ref.
\onlinecite{ullrich}). The imaginary part of the Kohn-Sham response
function (\ref{2.2.5}) thus has to be put in by hand. In the following,
we choose a value of $\eta = 0.1$ meV, corresponding to about 0.1-1\%
of the excitation energies under study.

Inserting everything into the response equation (\ref{2.2.4}), we obtain
\begin{eqnarray}\label{2.2.12}
\lefteqn{
 n_{1l\sigma}(r,\omega) =
\int\limits_0^\infty\!dr' \:r' \sum_{l'=-N}^N \chi_{ll' \sigma}(r,r',\omega)
\; v_{1l'\sigma}( r',\omega)} \nonumber\\
&+& 4\pi^2 {e^*}^2
\sum_\tau \int\limits_0^\infty\!dr'\:r'\int\limits_0^\infty\!dr''\:r''
\sum_{l'=-N}^N \chi_{ll' \sigma}(r,r',\omega) \:
n_{1l'\tau}(r'',\omega) I_{l'}(r',r'')
 \nonumber\\
&+& 2\pi \sum_\tau \int\limits_0^\infty\!dr'\:r'
 \sum_{l',m=-N}^N \chi_{ll' \sigma}(r,r',\omega) \:
  n_{1m\tau}(r',\omega) f_{{\rm xc},(l'-m)\sigma \tau} (r') \;.
\end{eqnarray}
Solving for the density response $n_1({\bf r},\omega)
= n_{1\uparrow}({\bf r},\omega)+n_{1\downarrow}({\bf r},\omega)$ 
for a given value of $\omega$
thus requires inversion of a complex matrix  of dimension 
$2N_{\rm grid}(2N+1) = 1400$ (for $N=3$ and $N_{\rm grid}=100$),
which poses no problem in practice.

\subsection{External potential}\label{sec2.3}
Let us now turn to the specific form of the bare confining potential of the
quantum dot, $v_{\rm dot}(r,\varphi)$, used to construct the electronic
ground state in Sec. \ref{sec2.1}. We first of all restrict ourselves to
considering only potentials that have inversion symmetry. In other words, 
the expansion (\ref{2.1.7}) of $v_{\rm dot}(r,\varphi)$ contains only 
angular components $v_{ {\rm dot},l}(r)$ with even $l$. We can then
replace $e^{il\varphi}$ in eq. (\ref{2.1.8}) by $\cos(l\varphi)$.
Furthermore, to reduce the computational effort we shall restrict the values 
of $l$ to $l=0, \pm2, \pm4$.

We first consider the circularly symmetric part of the confining potential,
$v_{ {\rm dot},0}(r)$. This component describes the degree of non-parabolicity
of our quantum dot. Pfannkuche and Gerhards \cite{pfannkuche} assume a form
\begin{equation}\label{2.3.1}
v_{ {\rm dot},0}(r) = \frac{m^*}{2}(\omega_0^2 \, r^2
+  \omega_1^2 \, r^4) \;,
\end{equation}
where $\omega_1 \ll \omega_0$. A different approach has been chosen by
Broido {\em et al.} \cite{broido}: they construct $v_{ {\rm dot},0}(r)$
as the electrostatic potential associated with a
two-dimensional jellium disk of radius $R$ and uniform positive areal
charge density $n_+$. Their result is
\begin{eqnarray}\label{2.3.2}
v_{ {\rm dot},0}(r) &=& v_0 - 
4{e^*}^2 n_+ \, R \, E\left(\frac{r}{R}\right),
\quad r<R\nonumber \hspace{2.0cm}\\
v_{ {\rm dot},0}(r) &=& v_0 - 4{e^*}^2 n_+ r\nonumber\\
\lefteqn{ \hspace{-0.5cm} \times \left[
E\left(\frac{R}{r}\right) - \left(1-\frac{R^2}{r^2}\right)
K\left(\frac{R}{r}\right)\right], \quad r>R\:,}
\end{eqnarray}
where $v_0 = 2 \pi {e^*}^2 n_+ \, R$. Here, $K$ and $E$ denote
complete elliptic integrals of the first and second kind.
It is easy to see that (\ref{2.3.1}) is the small-$r$ expansion of
(\ref{2.3.2}), identifying the coefficients as
\begin{eqnarray}\label{2.3.3}
\omega_0^2 &=& \frac{ \pi{e^*}^2 n_+}{m^* R} \nonumber\\
\omega_1^2 &=& \frac{3\pi{e^*}^2 n_+}{16m^*  R^3} \;.
\end{eqnarray}
The main difference between the two forms of $v_{ {\rm dot},0}(r)$ is that
(\ref{2.3.1}) grows as $r^4$ for large $r$, whereas (\ref{2.3.2}) approaches
the constant $v_0$ as $1/r$. 
This difference is of less importance for the electronic
ground state, since the two potentials are very similar in the interior region
of the dot where the electronic density is concentrated, but it can be 
expected to substantially affect the electronic excitations.

We now turn to the components of $v_{\rm dot}$ that break the
circular symmetry. In Refs. \onlinecite{gudmundsson,pfannkuche}
this is accomplished by including terms of square symmetry, i.e.
proportional to $x^2y^2$. Similarly, one can add on terms proportional to $x^2$
or $y^2$, describing elliptic elongation of the dot along the $x$ or $y$ axis.
\cite{hirose} In this manner, one arrives at
\begin{eqnarray}\label{2.3.4}
v_{ {\rm dot},\pm 2}(r) &=& a\, r^2 \nonumber\\
v_{ {\rm dot},\pm 4}(r) &=& b\, r^4 \;,
\end{eqnarray}
which introduces two more adjustable parameters $a$ and $b$, in addition to 
$\omega_0$ and $\omega_1$.

Again, an alternative approach to the construction of 
$v_{\rm dot}(r,\varphi)$ is
to start out with a flat jellium disk of uniform positive charge $n_+$,
but this time with a non-circular shape parametrized as $R(\varphi)$.
Once a particular form for $R(\varphi)$ has been chosen, the associated
electrostatic potential is calculated as
\begin{eqnarray}\label{2.3.5}
\lefteqn{
v_{\rm dot}(r,\varphi) =  \tilde{v}_0 } \nonumber\\ 
&-& {e^*}^2 n_+
\int\limits_0^{2\pi} d\varphi' \! \int\limits_0 ^{R(\varphi')}
\frac{r'\, dr'}{\sqrt{r^2 + r'^2 - 2rr' \cos(\varphi- \varphi')}} \;,
\end{eqnarray}
where now
\begin{equation}\label{2.3.6}
\tilde{v}_0 = {e^*}^2 n_+ \int_0^{2\pi} d\varphi' \: R(\varphi') \;.
\end{equation}
The $r'$-integral in (\ref{2.3.5}) can be performed analytically, and the
remaining integration over $\varphi'$ has to be done numerically for
general $R(\varphi')$. If $R(\varphi')= \mbox{const.}$, one recovers
the previous result (\ref{2.3.2}). The next step is then to construct the
angular components $v_{ {\rm dot},l}(r)$ using eq. (\ref{2.1.8}), which in
general requires a second straightforward numerical integration.

The large-$r$ behavior of the dot potential (\ref{2.3.5}) is found to be
\begin{equation}\label{2.3.7}
v_{\rm dot}(r\to \infty,\varphi) = \tilde{v}_0 - \frac{{e^*}^2 n_+}{2 r}
\int\limits_0^{2\pi} d\varphi' R(\varphi')^2 \;,
\end{equation}
independent of $\varphi$. This means that for large distances only
the $l=0$ component of $v_{\rm dot}(r,\varphi)$ survives (approaching the
constant $\tilde{v}_0$), and the higher-$l$ components go to zero. This 
is again in contrast with the form (\ref{2.3.4}) for the non-circular
components of the dot potential, which (unphysically) keep 
increasing with distance.
As noted before in the case of the circular dot, this difference is not 
expected to have a large impact on the electronic ground state, but it may 
become important for higher excitations. We therefore conclude that in
general it is preferable to work with dot potentials constructed
according to eq. (\ref{2.3.5}), thus avoiding effects caused by an
unphysical behavior in the large-$r$ region.

\section{Results and Discussion}\label{sec3}
The experiments presented in Refs. \onlinecite{fricke,lorke} were performed on
self-assembled InAs quantum dots embedded into GaAs. The diameter of the
dots is estimated to be about 20 nm and the height to be about 7 nm.
As outlined above, we treat the quantum dots as two-dimensional
systems. Within our model, we also ignore the presence of the wetting 
layer.

From their measurements, the authors of Refs. \onlinecite{fricke,lorke}
deduce an effective mass 
$m^* = 0.08\: m_e$ (where $m_e$ is the bare electronic mass). Furthermore,
we take the effective charge as $e^* = e/\sqrt{\varepsilon}$ ($e$ is the bare 
electronic charge), using $\varepsilon=15.15$
for the dielectric constant, i.e.  the bulk value of InAs,
and we employ an effective $g$-factor  $g^* = -0.44$.
For the curvature of the bare confining potential of the dot close to its
center, we take a value of
$\omega_0 = 45$ meV, which leads to $n_+ = 0.7\times 10^{15} \: {\rm cm}^{-2}$
via relation (\ref{2.3.3}) (for $R=100$ \AA).

The specific form of the  bare confining potential of the dot can now be
constructed using one of the two approaches discussed in Sec. {\ref{sec2.3}.
In the following, our choice is to construct $v_{\rm dot}(r,\varphi)$
as the electrostatic potential associated with a disk whose radius is 
parametrized as
\begin{equation}\label{3.1}
R(\varphi) = R_0 + R_2\sin^2\! \varphi + R_4 \sin^2 \!\varphi
\,\cos^2 \!\varphi \;.
\end{equation}
For $R_0$, we take the estimated radius of the dots, i.e. $R_0=100$ \AA.
The parameter $R_2$ indicates an elliptic elongation of the quantum dot
along the $y$-axis.
We adopt a value of $R_2 = 5$ \AA, as estimated In Ref. \onlinecite{fricke}.
For the parameter $R_4$ that causes an anisotropy of square symmetry,
no direct experimental numbers are available. In the following we choose
$R_4 = 83$ \AA, so that the value of $R_4 \sin^2 \!\varphi
\,\cos^2 \!\varphi$ is at most 20\% of $R_0$.
The resulting shape of the dot, a rectangle with
rounded-off corners, is shown as inset in Fig. \ref{figure1}. For comparison, 
a circle with radius $R_0$ is also indicated. 

In Fig. \ref{figure1} we plot the angular components $v_{ {\rm dot},l}(r)$ of 
the bare confining potential of the quantum disk parametrized by (\ref{3.1}). 
The top part shows the circularly symmetric part $v_{ {\rm dot},0}(r)$,
together with a parabolic potential that would correspond to the case
$v_{\rm dot}(r) = m^* \omega_0^2 r^2/2$.
The bottom part shows the $l=2$ and $l=4$ components.
Note that if the elongation of the dot is along the $x$-axis [replace
$R_2\sin^2 \varphi$ by $R_2\cos^2 \varphi$ in eq. (\ref{3.1})], then
$v_{ {\rm dot},2}(r)$ changes sign.
From Fig. \ref{figure1} it is evident that the deviation
from circular symmetry affects the confining potential mainly
in the region around the edge of the dot, whereas the inner region of the
dot sees a nearly parabolic potential. 

Let us now discuss our main numerical results. Fig. \ref{figure2} shows the
calculated peak positions of the photoabsorption cross section $\sigma(\omega)$
versus applied magnetic field, for a quantum dot with $N=2$ electrons. Here
and in the following, we assume the quantum dots to be elongated along the
$x$-axis. We then calculate the photoabsorption spectra for two
different polarizations of the FIR radiation, in $x$- and $y$-direction,
respectively. The symbols in  Fig. \ref{figure2} denote the average of the
two spectra. For comparison, the full lines show the expected behavior of
$\omega_\pm$ according to eq. (\ref{1.1}), with a fitted value of
$\omega_0 = 46.8$ meV. We see that for higher magnetic fields, the calculated
peaks follow the simple law of eq. (\ref{1.1}). For small $B$, however,
some deviations occur, and a splitting of about 2 meV remains even for
$B=0$ $T$, in accordance with observation. \cite{fricke,lorke}

For $N=2$, the quantum dot contains a full s-shell, \cite{note} and the
behavior is very similar to a parabolic dot. As soon as the p-shell gets
occupied, however, deviations from the parabolic case 
become much more pronounced.
In Fig. \ref{figure3} we plot the FIR peak positions for an $N=3$ quantum dot
versus $B$. The intensities of the absorption peaks are approximately 
indicated by the size of the symbols. The two top figures show results
for the non-circular dot, for two different polarizations of the FIR
radiation: parallel to the direction of elongation of the dot (the $x$-axis)
and perpendicular to it. In both cases we find very rich spectra. For small
magnetic fields, there are substantial differences between the two 
polarizations, for $B>6$ $T$ these differences disappear.

The bottom part of Fig. \ref{figure3} shows the spectra for a circular
(but still non-parabolic) dot. These results have been obtained by setting
$v_{{\rm dot},l}(r)=0$ for $l=\pm2,\pm4$, but using the same $v_{\rm dot,0}(r)$
as in the two figures above. By comparison between the lower and the two
upper parts of Fig. \ref{figure3}, we can now clearly distinguish between
those effects related to breaking the circular symmetry and those caused
by non-parabolicity. The latter leads to a splitting of the $\omega_+$ mode
into two almost equally strong subpeaks, $\omega_+^{(1)}$ and 
$\omega_+^{(2)}$, separated by about 5 meV, plus the appearance of a weaker
signal connected with the $\omega_-$ mode and approximately 10 meV below it.
We mention that similar results have been previously obtained by Hawrylak
and coworkers. \cite{hawrylak,wojs}

By introducing a non-circular anisotropy, the $\omega_+$ mode acquires a 
{\em third} subpeak $\omega_+^{(3)}$ in agreement with the experimentally
observed behavior (see Fig. 3 of Ref. \onlinecite{lorke}). We find that
the evolution of this signal with
magnetic field is different from that of the two other subpeaks of
$\omega_+$: it becomes weaker with increasing $B$ and its separation 
from $\omega_+^{(1)}$ and $\omega_+^{(2)}$ is growing.

To understand the origin of the three-fold
splitting of $\omega_+$, it is helpful to
resort to a simple single-particle picture. Fig. \ref{figure4} shows
the energy levels of a single electron confined in the bare potential of
our quantum dot at $B=12$ $T$, versus angular momentum quantum number
$l$ (see footnote \onlinecite{note}). The dashed lines connect energy levels
to which one can assign the same principal quantum number $n$. The
distribution of energy levels shown in Fig. \ref{figure4} is very similar to
that for a parabolic dot, which is governed by the formula
\begin{equation}\label{3.2}
E_{nl} = (2n+|l|+1)\sqrt{\omega_0^2 + \frac{\omega_c^2}{4}}
-\frac{\omega_c\,l}{2}
\end{equation}
(see the discussion in the review article by Ashoori\cite{ashoori}).
In contrast with the strictly parabolic case, the vertical distances between
the levels in Fig. \ref{figure4} are not constant.

In this simple picture (which remains qualitatively valid for $N>1$
even if the electron-electron interaction is included), the subpeaks
of $\omega_+$ can be identified with single-particle transitions. 
For circular symmetry, these are governed by the selection rule for dipole
transitions, $\Delta l = \pm 1$. In this simplified scenario, $\omega_+^{(1)}$
and $\omega_+^{(2)}$ arise from $\rm 1s \to 1p_-$ and $\rm 1p_+ \to 2s$
transitions, as indicated in Fig. \ref{figure4}.
Breaking the circular symmetry means that the selection rule
can be violated: we find that the $\omega_+^{(3)}$ mode originates from
the transition $\rm 1p_+ \to 1d_-$, i.e. $\Delta l = -3$.
The oscillator strength of this mode is
of course directly related to the degree of anisotropy, which in our case
is only small.

We have found that the position of $\omega_+^{(3)}$ with respect to 
$\omega_+^{(1)}$ and $\omega_+^{(2)}$ is insensitive to small variations in the
choice of the parameters $R_2$ and $R_4$ in eq. (\ref{3.1}). A weak signal at
$\omega_+^{(3)}$ will be present even if only one of the two is nonzero.
However, the value of $R_2=5$ {\AA} is more or less dictated by the 
experimentally observed splitting between $\omega_+$ and $\omega_-$ for 
$N=2$ at zero magnetic field, see above. One then observes that choosing
a finite value for $R_4$ increases the intensity of the third subpeak.

The central result of this work is presented in Fig. \ref{figure5}.
It shows a comparison between the calculated and measured \cite{fricke}
peak positions in the photoabsorption spectra for quantum dots with 
$1 \le N \le 6$ electrons at $B=12$ $T$. We find that the main experimental
features are reproduced by the calculation. For $N=1,2$, the system behaves
very similarly to a circular parabolic dot, as noted before, i.e. there are
only two signals at $\omega_+$ and $\omega_-$. As soon as the p-shell becomes
occupied, i.e. for $N\ge 3$, the $\omega_+$ mode splits up. As explained
above, the $\omega_+^{(3)}$ signal (indicated here by the open circles)
is related to a breaking of the circular symmetry. 

We note that the calculation yields a splitting between the three subpeaks of  
$\omega_+$ that is greater than the one found in the experiment. Also, 
for $N=6$ the experiment yields only a single signal at $\omega_+$. These 
differences between theory and experiment are to be attributed to the 
simplified nature of our model that treats the self-assembled dots as 
two-dimensional disks. In particular, the observed differences
at $N=6$ are most likely due to our neglecting the presence of the wetting
layer. 

The main effect of the wetting layer is to introduce
a continuum of states above a certain energy threshold \cite{hawrylak},
limiting the number of bound states localized in the self-assembled dot.
As the number
of electrons grows, the energy levels are shifted towards higher energies
due to the increasing interaction energy, and more and more states are pushed
into the continuum, up to a point where no additional electron can be bound.
From the absence of $\omega_+^{(2)}$ and $\omega_+^{(3)}$ for
$N=6$ observed in the experiment, we infer that
the relevant states involved in the transitions (2s and $\rm 1d_-$)
would in reality fall into the wetting layer continuum and the transitions
would lose most of their strength. However, 
for $N<6$ these states must still be located in the discrete part of the
energy spectrum, since the associated transitions are present in the experiment.

This effect has been accounted for in Ref. \onlinecite{wojs} by using a
truncated basis of only few bound states in the numerical diagonalization,
and it was indeed found that for $N=6$ there is only a single peak at
$\omega_+^{(1)}$. However, the calculations in Ref. \onlinecite{wojs} assumed
circular symmetry and produced
at most a twofold splitting of $\omega_+$ for $N=3,4,5$, in contrast with
experiment. We therefore conclude that, in spite of the deviations from
experiment mentioned above, our calculations clearly establish
that the presence of $\omega_+^{(3)}$ for $N=3,4,5$
is due to a breaking of circular symmetry of the quantum dots.

\section{conclusion}
In this work, we have developed a theoretical description of collective
charge-density excitations of non-circular quantum dots in a magnetic field. 
The computational scheme presented here allows one to obtain information
on the geometry of quantum dots from their electronic response properties.

Our specific aim
was to reproduce and explain recent experimental results, obtained for
self-assembled InAs quantum dots. In these FIR photoabsorption
measurements, one detects a threefold splitting of the upper branch $\omega_+$
of the collective charge-density mode in a magnetic field, 
depending on the number of electrons present in the dot.
Our study has shown that these experimental features are closely
related to the shape
of the dots: in addition to non-parabolicity (due to the finite
dot radius), it is essential to account for anisotropy effects, leading to 
a non-circular confining potential. To our knowledge, we have presented here
the first fully self-consistent spin-density-functional calculations of
ground state plus linear response for anisotropic quantum 
dots with up to 6 electrons.

With our calculations we were able to reproduce and explain the main features
of the FIR spectroscopy measurements conducted in Refs. 
\onlinecite{fricke,lorke}. However, the agreement was not fully quantitative.
This may be attributed to the simplicity of our model, which treats 
self-assembled, lens-shaped (with a possible elliptic or pyramidal distortion)
quantum dots sitting on a wetting layer as two-dimensional quantum disks.
It may safely be expected that this model leads to much better quantitative
results for quantum dot devices that are produced by mesa-etching techniques,
also known as vertical quantum dots \cite{tarucha}.

\acknowledgements

This work was supported by
Research Board Grant RB 96-071 from the University of Missouri and by
NSF grant No. DMR-9706788.  We thank Axel Lorke for stimulating discussions.



\begin{figure}
\unitlength1cm
\begin{picture}(15.0,17.0)
\put(-11.0,-17.0){\makebox(15.0,17.0){
\includegraphics{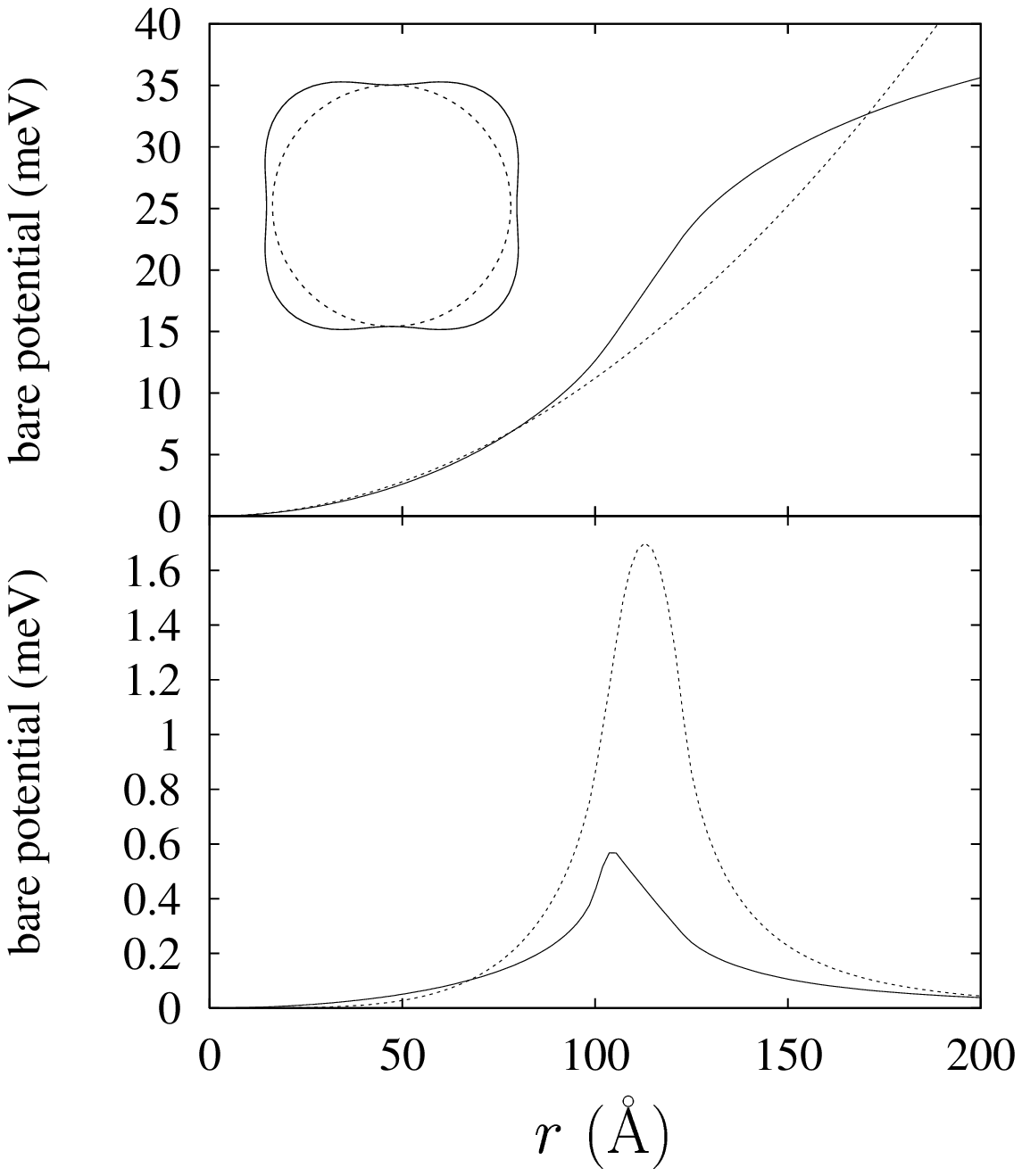}
}}
\end{picture}
\caption{
Angular components $v_{ {\rm dot},l}(r)$ of the bare confining potential
of a quantum disk parametrized by eq. (\ref{3.1}). The shape of the disk
is shown in the inset, together with a circle of radius $R_0=100$ \AA.
Top: $l=0$ component (full line). For comparison, the dashed line
shows a a parabolic potential with the same curvature in the interior region.
Bottom: $l=2$ (full line) and $l=4$ (dashed line) components.
}
\label{figure1}
\end{figure}

\begin{figure}
\unitlength1cm
\begin{picture}(15.0,17.0)
\put(-11.0,-18.5){\makebox(15.0,17.0){
\includegraphics{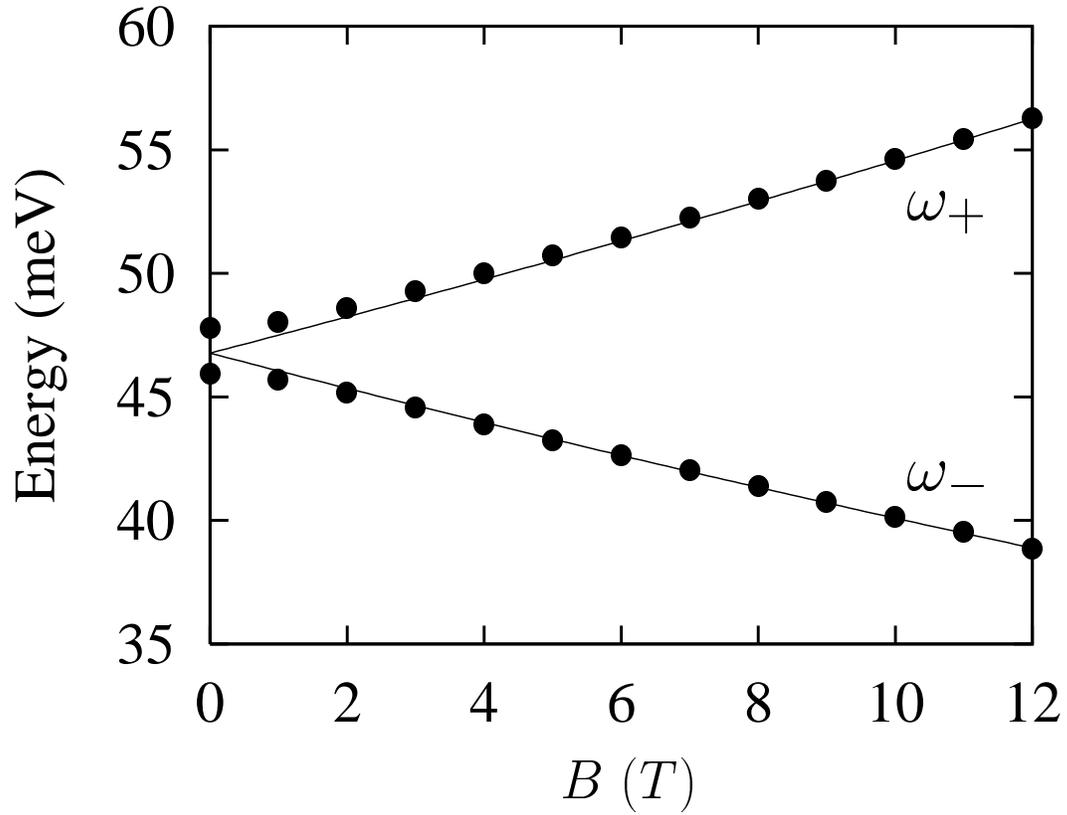}
}}
\end{picture}
\caption{
Calculated peak positions of the photoabsorption spectrum 
for a non-circular, non-parabolic quantum dot with $N=2$ electrons.
The full lines show the results one would obtain
for a circular parabolic dot, see eq. (\ref{1.1}).
}
\label{figure2}
\end{figure}

\begin{figure}
\unitlength1cm
\begin{picture}(15.0,20.5)
\put(-9.5,-12.5){\makebox(15.0,20.5){
\includegraphics{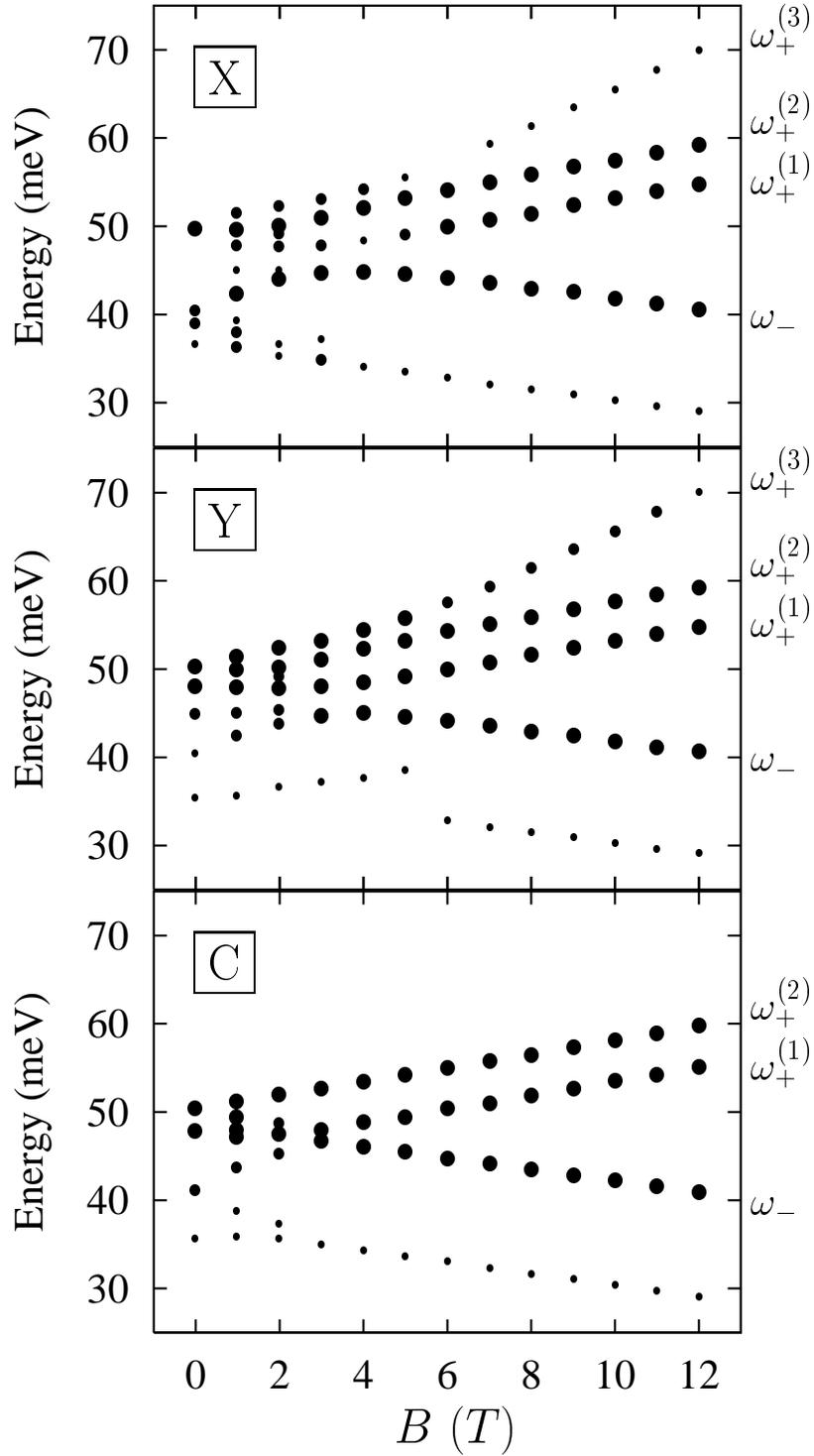}
}}
\end{picture}
\caption{
Calculated peak positions of the photoabsorption spectrum
for a quantum dot with $N=3$ electrons. Bottom: non-parabolic but circular
dot. Top and middle: non-parabolic, non-circular dot, with polarization 
of the FIR radiation parallel (X) and perpendicular (Y) to the direction
of elongation. The $\omega_+$ mode splits up into three subpeaks,
$\omega_+^{(1)}$, $\omega_+^{(2)}$ and $\omega_+^{(3)}$, as indicated.
}
\label{figure3}
\end{figure}

\begin{figure}
\unitlength1cm
\begin{picture}(15.0,17.0)
\put(-11.0,-15.0){\makebox(15.0,17.0){
\includegraphics{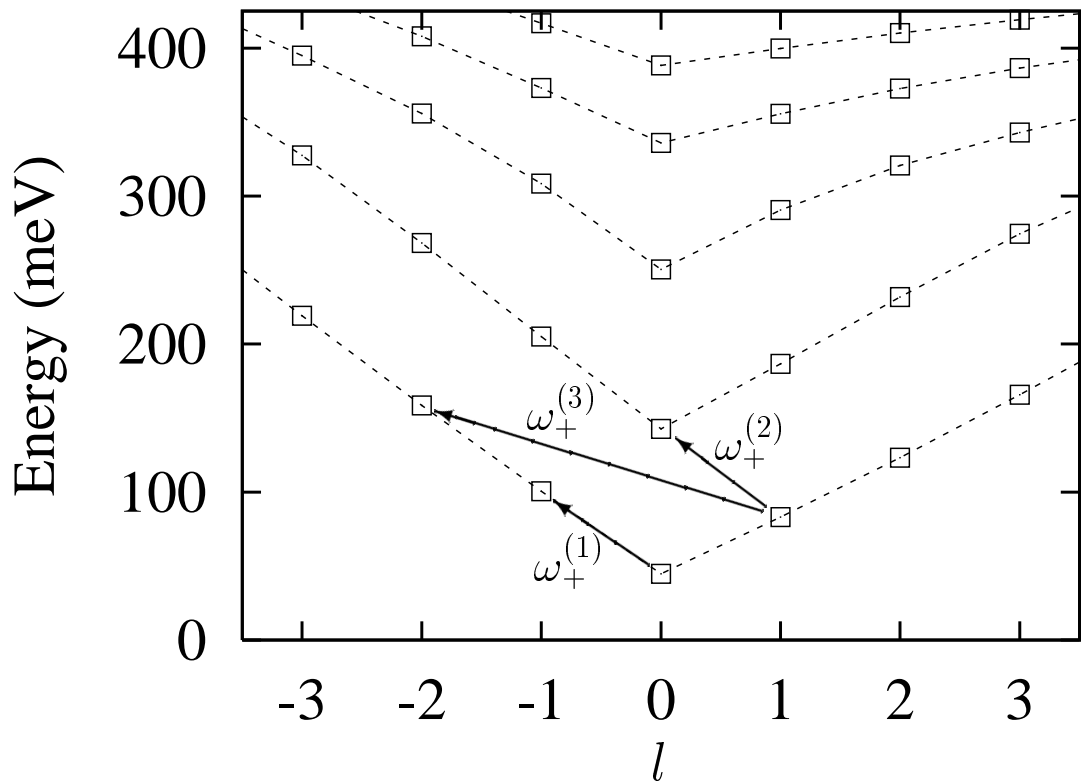}
}}
\end{picture}
\caption{
Energy levels of a single electron in the bare potential of the
quantum dot from Fig. \ref{figure1} at $B=12$ $T$. The levels are drawn
versus their angular momentum quantum number $l$, and dashed lines connect
levels with the same principal quantum number $n$.  The arrows indicate
the single-particle transitions that constitute the three-fold splitting
of $\omega_+$.
}
\label{figure4}
\end{figure}

\begin{figure}
\unitlength1cm
\begin{picture}(15.0,19.0)
\put(-9.5,-14.0){\makebox(15.0,19.0){
\includegraphics{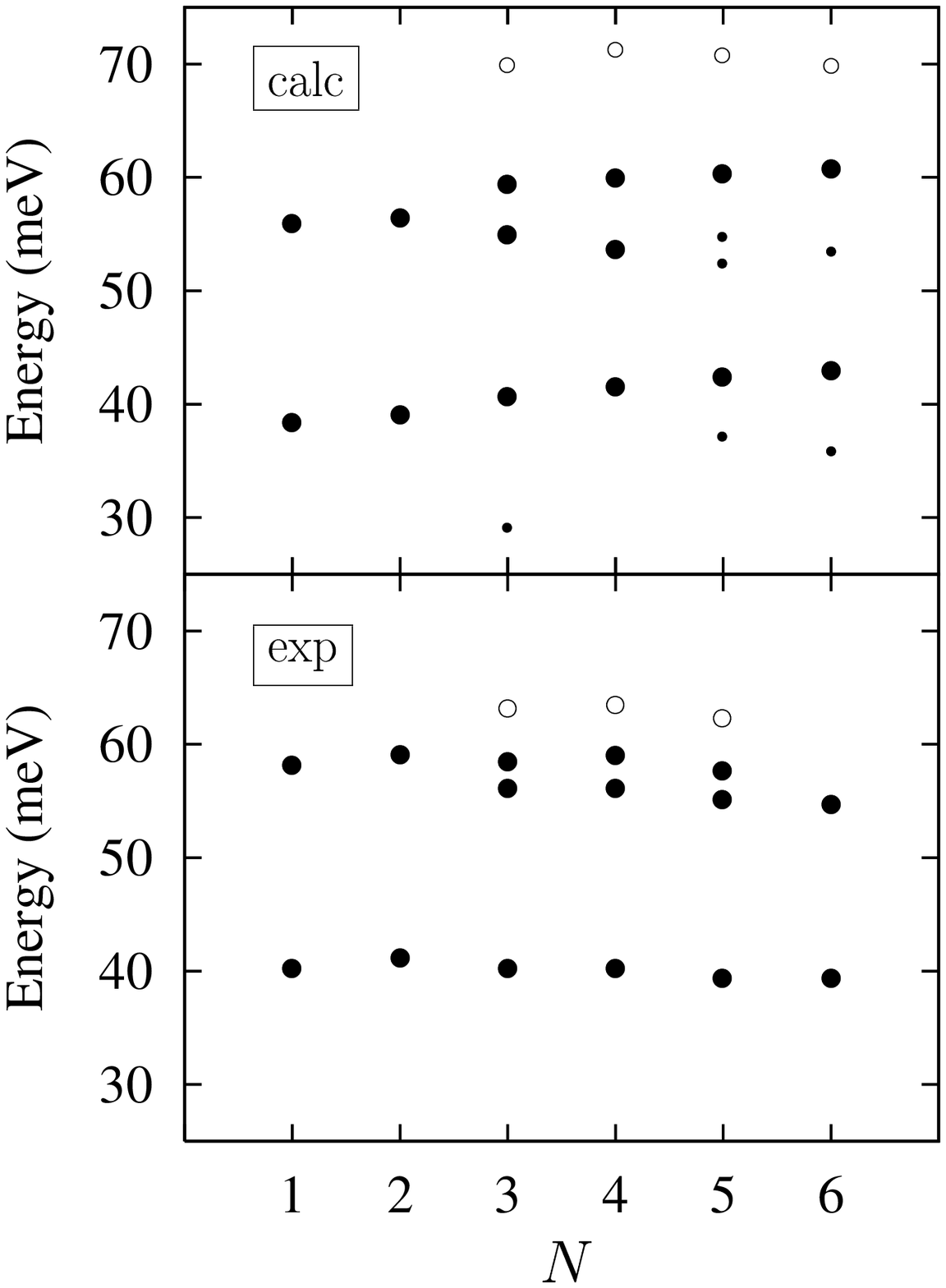}
}}
\end{picture}
\caption{
Comparison between the calculated and experimental
peak positions in the photoabsorption spectra for quantum dots with
$1 \le N \le 6$ electrons at $B=12$ $T$. The open circles indicate those
signals that we find to be related to a broken circular symmetry of the dots.
}
\label{figure5}
\end{figure}


\begin{references}
\bibitem{note}
For a system with broken circular symmetry, the angular momentum is no longer
a good quantum number, and the electronic orbitals become linear combinations
of pure-$l$ states, see eq. (\ref{2.1.4}). However, if the anisotropy of the
dots is viewed as a small perturbation, one may continue to speak of 
s,p,d,... electrons, since for each $\psi_{j\sigma}$ the expansion (\ref{2.1.4})
is then dominated by a single term with well-defined $l$ (we use the
standard notation s for $l=0$, $\rm p_\pm$ for $l=\pm1$ etc.).
\end{references}
\end{document}